\documentclass [12pt]{article}
\usepackage{graphicx}

\begin{document}

\begin{center}
\textbf{The vector and tensor asymmetries and deuteron wave function for
different nucleon-nucleon potentials}
\end{center}

\begin{center}
\textbf{V. I. Zhaba}
\end{center}

\begin{center}
Uzhgorod National University, Department of Theoretical Physics
\end{center}

\begin{center}
54, Voloshyna St., Uzhgorod, UA-88000, Ukraine
\end{center}

\begin{center}
E-mail address: viktorzh@meta.ua
\end{center}

\textbf{Abstract}

Within the framework of the study of radiative corrections to the
polarization observed in elastic ed- scattering in leptonic variables have
been calculated the Born values of vector $A_B^L $, $A_B^T $ and tensor
$A_B^{LL} $, $A_B^{TT} $, $A_B^{LT} $ asymmetries. The deuteron wave
functions in coordinate representation for eight nucleon-nucleon potentials
(Nijm1, Nijm2, Nijm93, Reid93, Argonne v18, OBEPC, MT and Paris) have been
used for numerical calculations of vector and tensor asymmetries. The
angular-momentum dependence of values vector $A_B^i (p,\theta )$ and tensor
$A_B^{ij} (p,\theta )$ asymmetries have been also evaluated in 3D format for
Reid93 potentials.


\textbf{1. Introduction}

Deuteron can be used as a target for an electron beam or a
particle that is scattered on protons and nuclei. For example, in
\cite{1} the results for spin-dependent scattering of electrons on
polarized protons and deuterons for the BLAST experiment conducted
in MIT-Bates are given. On the electron-positron storage VEPP-3
\cite{2} obtained the results of measuring the components of the
analyzing powers $T_{2i}$ in the photodisintegration reaction of a
tensor-polarized deuteron.

The spin observed in dp- scattering and the T-invariance test in
the application of the modified Glauber theory was investigated in
paper \cite{3}. A complete set of deuteron analyzing powers in an
elastic dp- scattering at 190 MeV/nucleon is indicated in
\cite{4}. The proton and deuteron analyzing powers and 10 spin
correlation coefficients were measured in \cite{5} for elastic ð+d
scattering with the energy of bombarding protons 135 and 200 MeV.
The polarization observables for $^{2}$H(d,p)$^{3}$H and
$^{2}$H(d,n)$^{3}$He reactions for five potentials and their
comparison with experimental data's are discussed in \cite{6}.

In \cite{7} the angular dependence of the tensor $À_{óó}$ and the
vector $À_{ó}$ analyzing powers in inelastic (d,d')- scattering of
deuterons with a momentum at 9.0 GeV/c on hydrogen and carbon is
measured. Further research and prospects for the process (d,d')
are analyzed in \cite{8}.

For the theoretical study of the mechanisms and characteristics for these
processes with the participation of a deuteron, it is necessary to know
exactly the deuteron wave function in the coordinate or momentum
representation, as well as the deuteron form-factors.

In the last detailed review \cite{9}, the static parameters of
deuteron obtained from DWF for different nucleon-nucleon
potentials and models are systematized, as well as an overview,
list and characteristics for analytic forms of DWF in the
coordinate representation are given.

In this paper we use the analytic forms of DWF for theoretical calculations
of a set of tensor and vector asymmetries. Nucleon-nucleon realistic
phenomenological potentials of Nijmegen group (NijmI, NijmII, Nijm93,
Reid93) and Argonne group (Argonne v18) as well as other widely used and
popular potentials (OBEPC, MT, Paris) are used for numerical calculations.

\textbf{2. DWF in coordinate representation}

Among the various numbers of DWF parameterization forms in the coordinate
representation, one can distinguish the following:

1) the parameterization, which was applied to Paris potential
\cite{10}:

\begin{equation}
\label{eq1}
\left\{ {\begin{array}{l}
 u\left( r \right) = \sum\limits_{j = 1}^N {C_j \exp \left( { - m_j r}
\right),} \\
 w\left( r \right) = \sum\limits_{j = 1}^N {D_j \exp \left( { - m_j r}
\right)\left[ {1 + \frac{3}{m_j r} + \frac{3}{\left( {m_j r} \right)^2}}
\right];} \\
 \end{array}} \right.
\end{equation}

where $m_j = \beta + (j - 1)m_0 $; $\beta = \sqrt {ME_d } $, $m_{0}$=0.9
fm$^{ - 1}$; $M$ -- nucleon mass; $E_{d}$ -- binding energy of the deuteron;

2) the parameterization for Moscow potential \cite{11}:

\begin{equation}
\label{eq2}
\left\{ {\begin{array}{l}
 u(r) = r\sum\limits_{i = 1}^N {A_i \exp ( - a_i r^2),} \\
 w(r) = r^3\sum\limits_{i = 1}^N {B_i \exp ( - b_i r^2);} \\
 \end{array}} \right.
\end{equation}

3) the analytical form for Nijmegen group potentials (NijmI,
NijmII and Nijm93) \cite{12}:

\begin{equation}
\label{eq3}
\left\{ {\begin{array}{l}
 u(r) = r^{3 / 2}\sum\limits_{i = 1}^N {A_i \exp ( - a_i r^3),} \\
 w(r) = r\sum\limits_{i = 1}^N {B_i \exp ( - b_i r^3).} \\
 \end{array}} \right.
\end{equation}

In this paper we use the DWF in the form (\ref{eq3}) for the
potentials of the Nijmegen group (NijmI, NijmII, Nijm93, Reid93)
and the Argonne group (Argonne v18). The coefficients of DWF for
these potentials are given in \cite{12,13}.

The search for coefficients of the analytical form (\ref{eq1}) was
made for the Bonn (OBEPC) potential \cite{14} and MT model
\cite{15}.

\textbf{3. The vector }$A_B^L $\textbf{, }$A_B^T $\textbf{ and tensor
}$A_B^{LL} $\textbf{, }$A_B^{TT} $\textbf{, }$A_B^{LT} $\textbf{
asymmetries}

In recent years, the problem of the study of leptonic radiative
corrections to elastic deuteron-electron scattering remains an
urgent issue \cite{16}.

To get the radiative corrections to the polarization observables
for the reaction $e^ - (k_1 ) + d(p_1 ) \to e^ - (k_2 ) + d(p_2
)$, it is necessary to parameterize the polarization state of the
target in terms of the 4-momenta of particles in this reaction
\cite{17}. The polarization state of the target describes the
deuteron polarization 4-vector $s_{\mu }$ and the quadrupole
polarization tensor $p_{\mu \nu }$. Such a parameterization for
the polarization state depends on the directions in which the
longitudinal and transverse components of the deuteron
polarization in rest frame are determined. The value $s_{\mu }$
describing the deuteron vector polarization.

The longitudinal $s^{(L)}$ and transverse $s^{(T)}$ polarization
4-vectors are defined as \cite{17}

\[
s_\mu ^{(T)} = \frac{(4\tau + \rho )k_{1\mu } - (1 + 2\tau )q_\mu - (2 -
\rho )p_{1\mu } }{\sqrt {Vc(4\tau + \rho )} };
\]

\[
s_\mu ^{(L)} = \frac{2\tau q_\mu - \rho p_{1\mu } }{M_D \sqrt {\rho (4\tau +
\rho )} }.
\]

Five Born values of vector $A_B^L $, $A_B^T $ and tensor $A_B^{LL}
$, $A_B^{TT} $, $A_B^{LT} $ asymmetries were considered For the
research problem and the search for ``radiative corrections to
polarization observables in elastic electron-deuteron scattering
in leptonic variables'' \cite{17}.

The spin-dependent parts of the cross-section is determined by the
vector polarization of the initial deuteron and longitudinal
polarization of the electron beam \cite{17,18}

\[
\frac{d\sigma _B^L }{dp^2} = - \frac{\pi \alpha ^2}{4\tau V^2}\frac{2 - \rho
}{\rho }\sqrt {\rho (4\tau + \rho )} G_M^2 ;
\]

\[
\frac{d\sigma _B^T }{dp^2} = - \frac{\pi \alpha ^2}{Vp^2}\sqrt {\frac{(4\tau
+ \rho )c}{\tau }} G_M G;
\]

where

\[
G = 2G_C + \frac{2}{3}\eta G_Q ;
\quad
c = 1 - \rho - \rho \tau ;
\quad
\eta = \frac{P^2}{4M_D^2 };
\quad
\rho = \frac{p^2}{V};
\quad
\tau = \frac{M_D^2 }{V}.
\]

In the laboratory system these expressions for cross-section lead
to the values of asymmetries (or the spin correlation
coefficients) in the elastic electron-deuteron scattering in the
Born approximation \cite{18}

\begin{equation}
\label{eq4}
\frac{d\sigma _B^L }{dp^2} = \frac{\pi }{\varepsilon _2^2 }\eta \sigma _{NS}
\sqrt {(1 + \eta )\left( {1 + \eta \sin ^2\left( {\frac{\theta _e }{2}}
\right)} \right)} \mbox{tg}\left( {\frac{\theta _e }{2}} \right)\sec \left(
{\frac{\theta _e }{2}} \right)G_M^2 ;
\end{equation}

\begin{equation}
\label{eq5}
\frac{d\sigma _B^T }{dp^2} = 2\frac{\pi }{\varepsilon _2^2 }\sigma _{NS}
\sqrt {\eta (1 + \eta )} \mbox{tg}\left( {\frac{\theta _e }{2}} \right)G_M
\left( {G_C + \frac{\eta }{3}G_Q } \right);
\end{equation}

where $\varepsilon _{2}$ are the scattered electron energy;
$G_{C}(p)$, $G_{Q}(p)$, $G_{M}(p)$ are deuteron form factors; \textit{$\theta $}$_{å}$ -- the scattering
angle of electron.

These asymmetries are formed thanks to the vector polarization of
the deuteron target (according to the longitudinal and transverse
direction of the spin 4-vectors) and the longitudinal polarization
of the electron beam \cite{17}

\begin{equation}
\label{eq6}
A_B^L = - \eta \sqrt {(1 + \eta )\left( {1 + \eta \sin ^2\left(
{\frac{\theta _e }{2}} \right)} \right)} \mbox{tg}\left( {\frac{\theta _e
}{2}} \right)\sec \left( {\frac{\theta _e }{2}} \right)G_M^2 I_0^{ - 1} ;
\end{equation}

\begin{equation}
\label{eq7}
A_B^T = - 2\sqrt {\eta (1 + \eta )} \mbox{tg}\left( {\frac{\theta _e }{2}}
\right)G_M \left( {G_C + \frac{\eta }{3}G_Q } \right)I_0^{ - 1} ;
\end{equation}

where

\[
I_0 = A + B\mbox{tg}^2\left( {\frac{\theta _e }{2}} \right).
\]

The ratio of the vector longitudinal and transverse polarization
asymmetries to the transverse can be written as \cite{18}

\begin{equation}
\label{eq8}
\frac{A_B^L }{A_B^T } = \frac{d\sigma _B^L / d\sigma _B }{d\sigma _B^T /
d\sigma _B } = \frac{2 - \rho }{4}\sqrt {\frac{\rho }{c\tau }} \frac{G_M
}{G}
\end{equation}

or in the laboratory system \cite{17}

\begin{equation}
\label{eq9}
\frac{A_B^L }{A_B^T } = \sqrt {\eta \left( {1 + \eta \sin ^2\left(
{\frac{\theta _e }{2}} \right)} \right)} \sec \left( {\frac{\theta _e }{2}}
\right)\frac{G_M }{G}.
\end{equation}

Four-vectors for tensor polarized deuteron target written as

$s_\mu ^{(I)} = \frac{2\varepsilon _{\mu \lambda \rho \sigma } p_{1\lambda }
k_{1\rho } k_{2\sigma } }{V\sqrt {Vc\rho } }$at $I=L$,$T$,$N$.

The part of the cross-section in the Born approximation depends on
the tensor polarization of the deuteron target \cite{17,18}

\begin{equation}
\label{eq10}
\frac{d\sigma _B^p }{dp^2} = \frac{d\sigma _B^{LL} }{dp^2}R_{LL} +
\frac{d\sigma _B^{TT} }{dp^2}(R_{TT} - R_{NN} ) + \frac{d\sigma _B^{LT}
}{dp^2}R_{LT} ,
\end{equation}

where three components for this cross-section:

\begin{equation}
\label{eq11}
\frac{d\sigma _B^{LL} }{dp^2} = \frac{\pi \alpha ^2}{p^4}2c\eta \left\{
{8G_C G_Q + \frac{8}{3}\eta G_Q^2 + \frac{2c + 4\tau \rho + \rho
^2}{2c}G_M^2 } \right\};
\end{equation}

\begin{equation}
\label{eq12}
\frac{d\sigma _B^{TT} }{dp^2} = \frac{\pi \alpha ^2}{p^4}2c\eta G_M^2 ;
\end{equation}

\begin{equation}
\label{eq13}
\frac{d\sigma _B^{LT} }{dp^2} = - \frac{\pi \alpha ^2}{p^4}4\eta (2 - \rho
)\sqrt {\frac{c\rho }{\tau }} G_Q G_M .
\end{equation}

In the laboratory system these parts lead to the following three
asymmetries (or analyzing powers) in the elastic electron-deuteron
scattering, that were induced by tensor polarization of the
deuteron target and an unpolarized electron beam \cite{18}

\[
\frac{d\sigma _B^p }{dp^2} = \frac{\pi }{\varepsilon _2^2 }\sigma _{NS}
\left[ {S_{LL} R_{LL} + S_{TT} (R_{TT} - R_{NN} ) + S_{LT} R_{LT} } \right]
\]

or this formula it is represented in \cite{17} as

\[
I_0 A_B^p = A_B^{LL} R_{LL} + A_B^{TT} (R_{TT} - R_{NN} ) + A_B^{LT} R_{LT}
,
\]

where $A_B^{LL} $, $A_B^{TT} $, $A_B^{LT} $ are the tensor polarizations
asymmetries:

\begin{equation}
\label{eq14}
A_B^{LL} = \frac{1}{2}\left\{ {8\eta G_C G_Q + \frac{8}{3}\eta ^2G_Q^2 +
\eta \left[ {1 + 2(1 + \eta )\mbox{tg}^2\left( {\frac{\theta _e }{2}}
\right)} \right]G_M^2 } \right\}I_0^{ - 1} ;
\end{equation}

\begin{equation}
\label{eq15}
A_B^{TT} = \frac{1}{2}\eta G_M^2 I_0^{ - 1} ;
\end{equation}

\begin{equation}
\label{eq16}
A_B^{LT} = - 4\eta \sqrt {\eta + \eta ^2\sin ^2\left( {\frac{\theta _e }{2}}
\right)} \sec \left( {\frac{\theta _e }{2}} \right)G_Q G_M I_0^{ - 1} .
\end{equation}

Between components there is a following connection $\frac{\pi }{\varepsilon
_2^2 }\sigma _{NS} S_{ij} = \frac{A_B^{ij} }{I_0 }$.

\textbf{4. Calculations and conclusions}

Figs. 1-5 illustrate the values of vector $A_B^L $, $A_B^T $ and tensor
$A_B^{LL} $, $A_B^{TT} $, $A_B^{LT} $ asymmetries (more precisely, it will
be an angle dependence or asymmetry for values $A_B^i $, $A_B^{ij} )$. The
calculations were made for Nijm1, Nijm2, Nijm93, Reid93, Argonne v18, OBEPC,
MT and Paris potentials. The angular dependence of vector $A_B^i $ and
tensor $A_B^{ij} $ asymmetries is clearly expressed and strongly manifested
for all these potentials.

The angular-momentum dependence of values vector $A_B^i (p,\theta
)$ and tensor $A_B^{ij} (p,\theta )$ asymmetries in 3D format for
Reid93 potential are given in Figs. 6-10. Here for tensor
asymmetries $A_B^{LL} $ and $A_B^{TT} $ there is a hump (peak) at
3.5-4 fm$^{-1}$ in the range of angles 0-180 degrees. For the
tensor asymmetry $A_B^{LT} $, on the contrary, there is a pit.

The obtained coefficients of vector and tensor asymmetries can be
applied for comparison with two sets of components for
cross-sections \cite{17}

$\frac{d\sigma _B^\beta }{dp^2} = V_{\beta A} ( - \theta
)\frac{d\sigma _B^A }{dp^2}$ at $A=L$, $T$ and \textit{$\beta
$}=$l$, $t$; (17)

$\frac{d\sigma _B^\beta }{dp^2} = T_{\beta A} ( - \theta
)\frac{d\sigma _B^A }{dp^2}$ at $A$=\textit{LL}, \textit{TT},
\textit{LT} and \textit{$\beta $}=\textit{ll}, \textit{tt},
\textit{lt }(18)

for polarization four-vectors

\[
s_\mu ^{(l)} = \frac{2\tau k_{1\mu } - p_{1\mu } }{M_D };
\quad
s_\mu ^{(n)} = s_\mu ^{(N)} ;
\quad
s_\mu ^{(t)} = \frac{k_{2\mu } - (1 - \rho - 2\rho \tau )k_{1\mu } - \rho
p_{1\mu } }{\sqrt {Vc\rho } }.
\]

Moreover, it is of interest to conduct the further study of
polarization observables in elastic lepton-deuteron scattering
including the lepton mass \cite{19} (and for the case of spin
correlation coefficients in the limit of zero lepton mass).

\pdfximage width 110mm {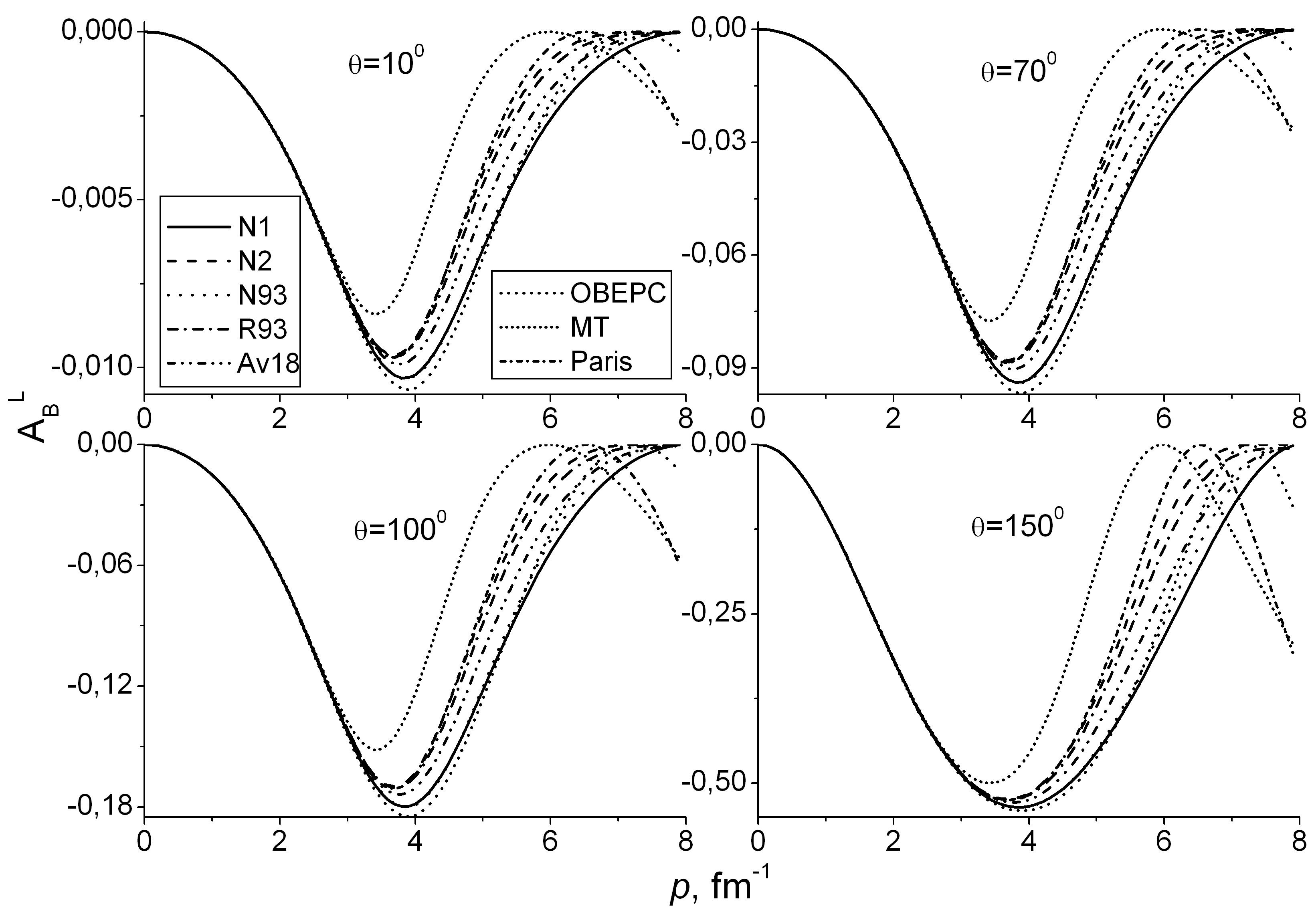}\pdfrefximage\pdflastximage

Fig.~1. The angular dependence of vector asymmetry $A_B^L $

\pdfximage width 110mm {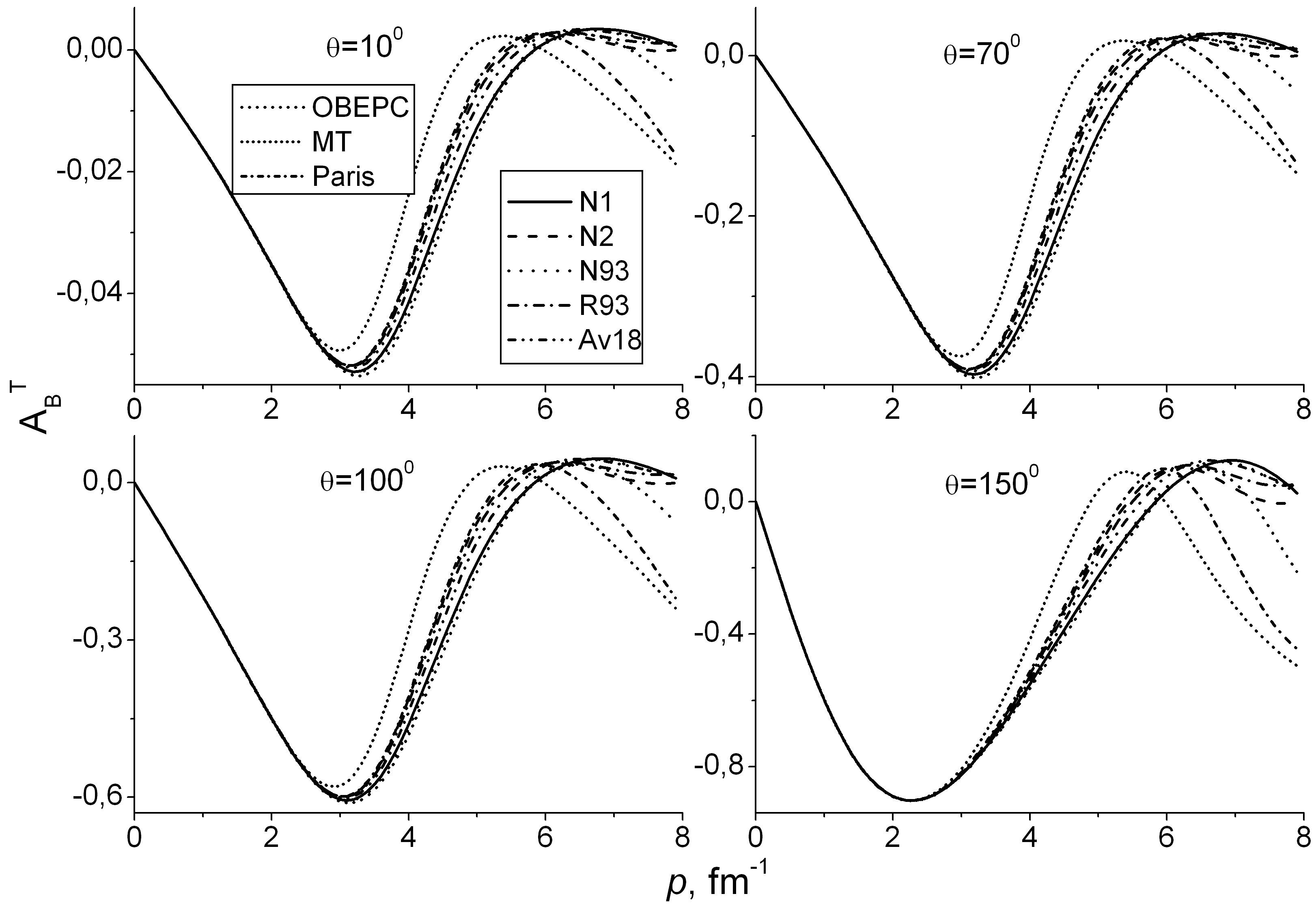}\pdfrefximage\pdflastximage

Fig.~2. The angular dependence of vector asymmetry $A_B^T $

\pdfximage width 110mm {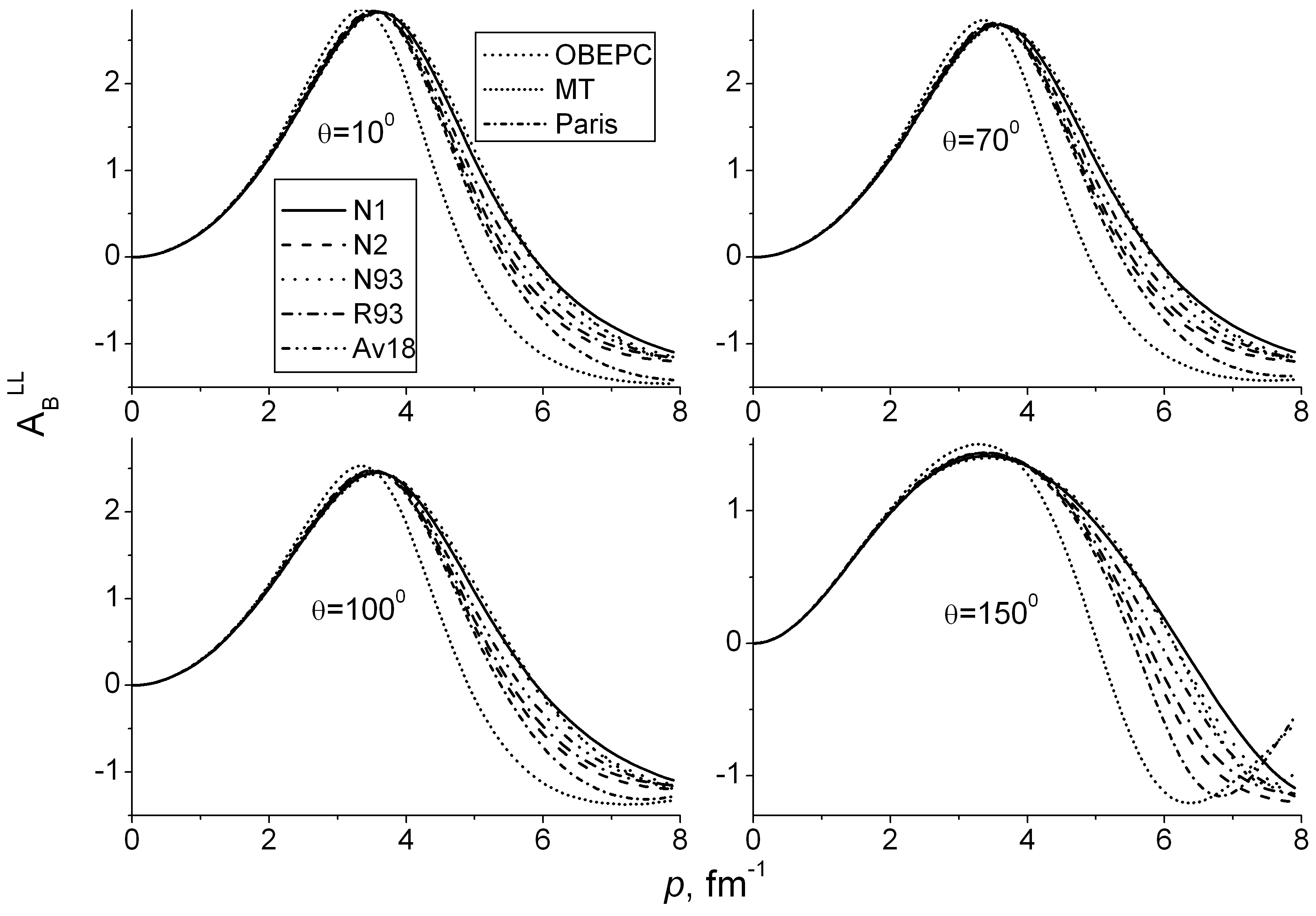}\pdfrefximage\pdflastximage

Fig.~3. The angular dependence of tensor asymmetry $A_B^{LL} $

\pdfximage width 110mm {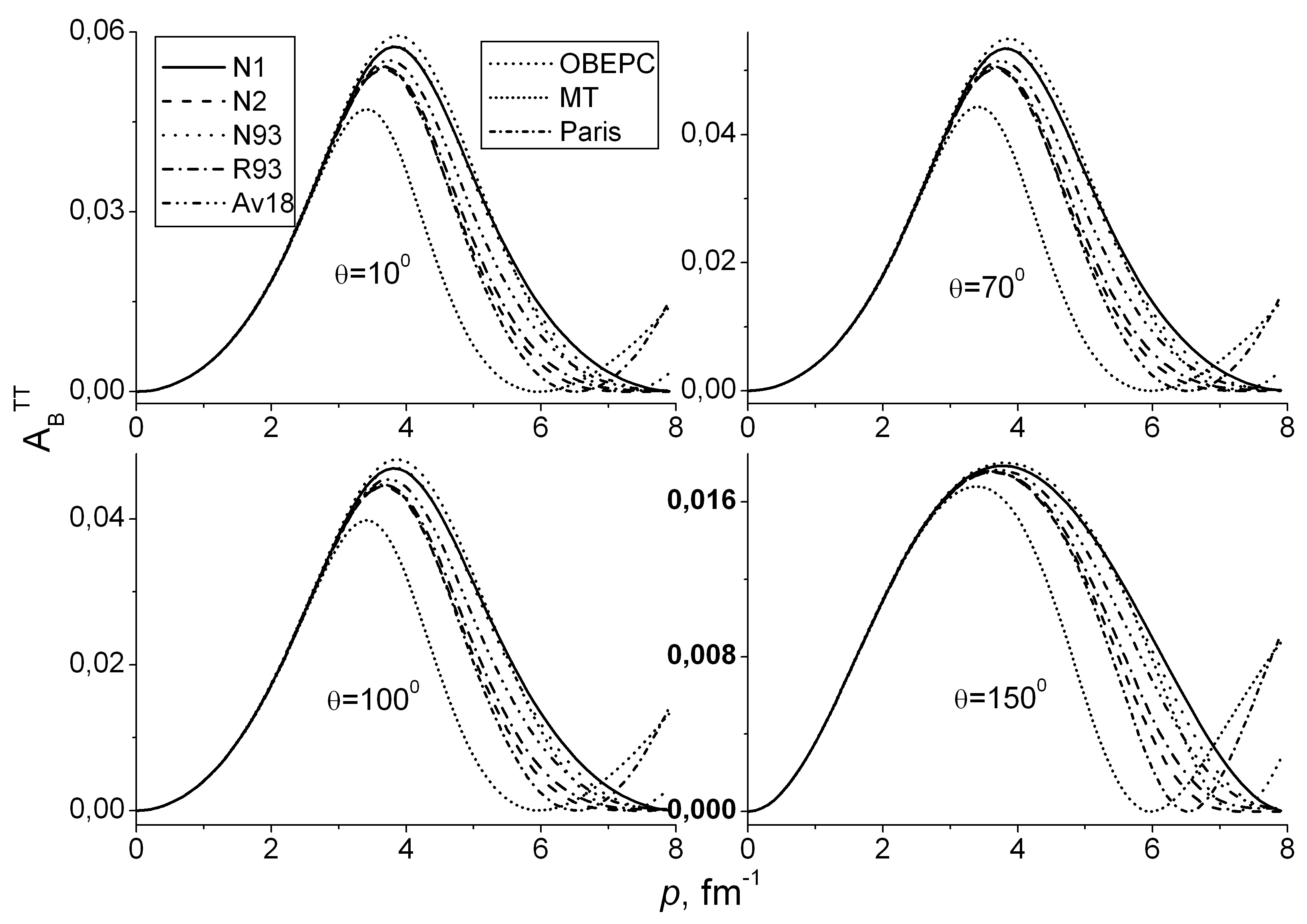}\pdfrefximage\pdflastximage

Fig.~4.  The angular dependence of tensor asymmetry $A_B^{TT} $

\pdfximage width 110mm {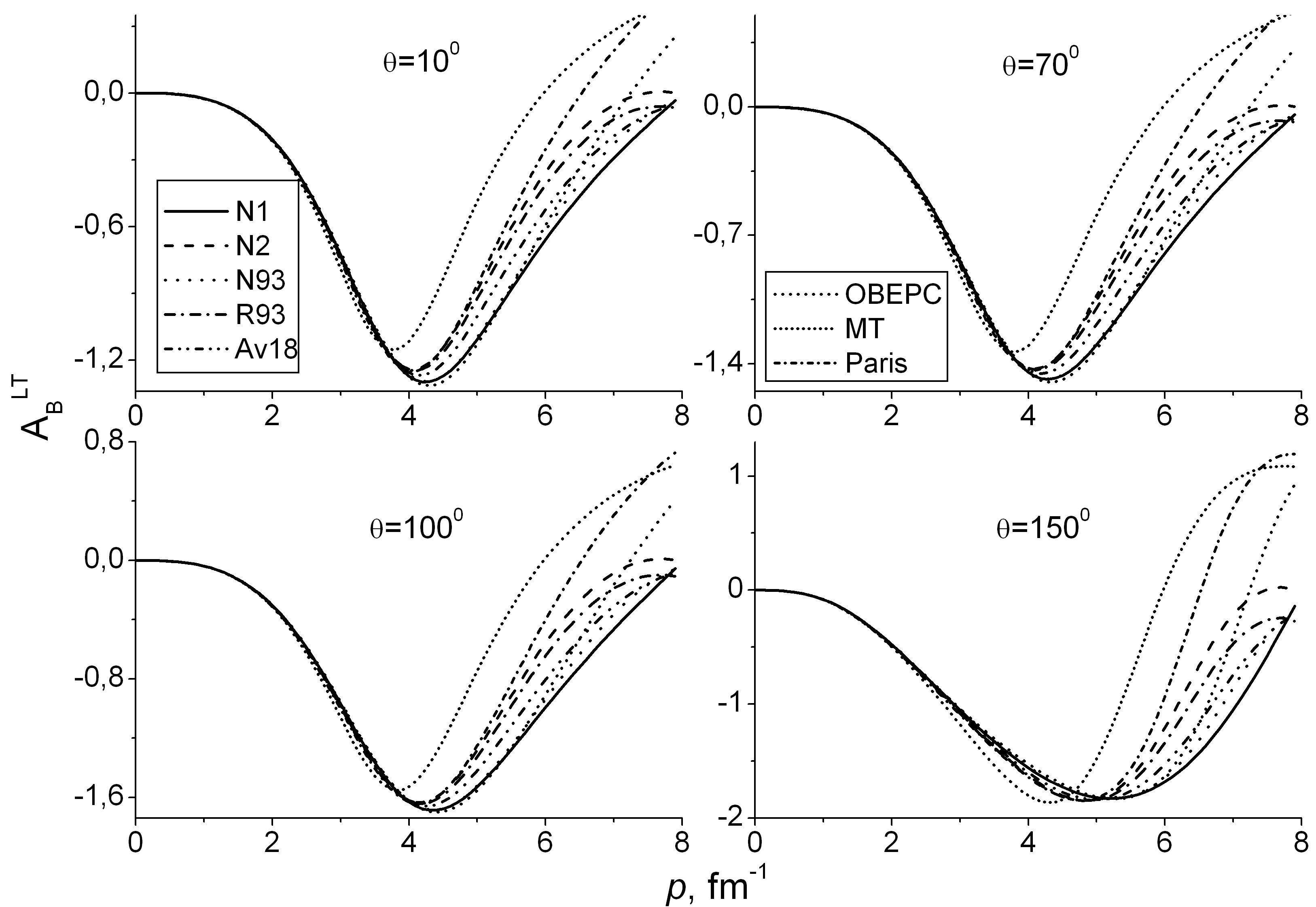}\pdfrefximage\pdflastximage

Fig.~5. The angular dependence of tensor asymmetry $A_B^{LT} $

\pdfximage width 90mm {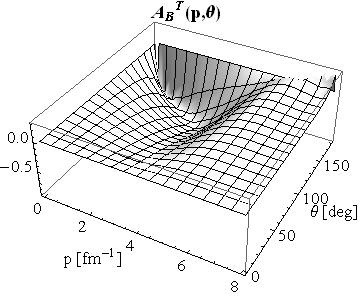}\pdfrefximage\pdflastximage

Fig.~6. The vector asymmetry $A_B^T $ for Reid93 potential

\pdfximage width 90mm {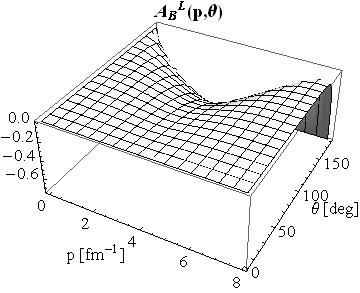}\pdfrefximage\pdflastximage

Fig.~7. The vector asymmetry $A_B^L $ for Reid93 potential

\pdfximage width 90mm {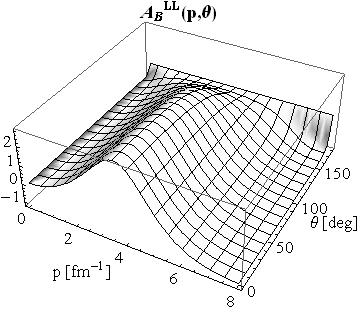}\pdfrefximage\pdflastximage

Fig.~8. The tensor asymmetry $A_B^{LL} $ for Reid93 potential

\pdfximage width 90mm {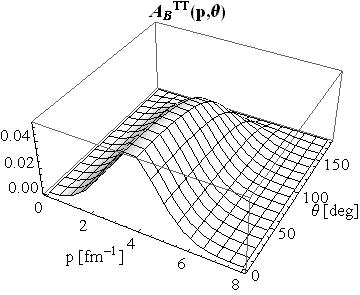}\pdfrefximage\pdflastximage

Fig.~9. The tensor asymmetry $A_B^{TT} $ for Reid93 potential

\pdfximage width 90mm {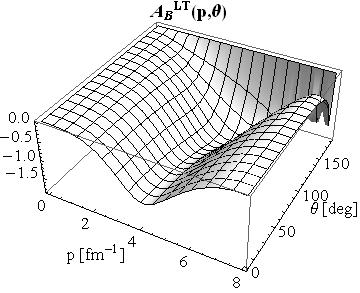}\pdfrefximage\pdflastximage

Fig.~10. The tensor asymmetry $A_B^{LT} $ for Reid93 potential

\end{document}